# Shannon information storage in noisy phase-modulated fringes and fringe-data compression by phase-shifting algorithms

**MANUEL SERVIN,**[*] **AND MOISES PADILLA**

*Centro de Investigaciones en Optica A. C., Loma del Bosque 115, 37150 Leon Guanajuato, Mexico.*
*[*]mservin@cio.mx*
*www.cio.mx*

**Abstract:** Optical phase-modulated fringe-patterns are usually digitized with *X*x*Y* pixels and 8 bits/pixel (or higher) gray-levels. The digitized 8 bits/pixel are raw-data bits, not Shannon information bits. Here we show that noisy fringe-patterns store much less Shannon information than the capacity of the digitizing camera. This means that high signal-to-noise ratio (*S/N*) cameras may waste to noise most bits/pixel. For example one would not use smartphone cameras for high quality phase-metrology, because of their lower (*S/N*) images. However smartphones digitize high-resolution (12 megapixel) images, and as we show here, the information storage of an image depends on its bandwidth and its (*S/N*). The standard formalism for measuring information are the Shannon-entropy *H*, and the Shannon capacity theorem (SCT). According to SCT, low (*S/N*) images may be compensated with a larger fringe-bandwidth to obtain high-information phase measurements. So broad bandwidth fringes may give high quality phase, in spite of digitizing low (*S/N*) fringe images. Most real-life images are redundant, they have smooth zones where the pixel-value do not change much, and data compression algorithms are paramount for image transmission/storage. Shannon's capacity theorem is used to gauge competing image compression algorithms. Here we show that phase-modulated phase-shifted fringes are highly correlated, and as a consequence, phase-shifting algorithms (PSAs) may be used as fringe-data compressors. Therefore a PSA may compress a large number of phase-shifted fringes into a single complex-valued image. This is important in spaceborne optical/RADAR phase-telemetry where downlink is severely limited by huge distance and low-power downlink. That is, instead of transmitting *M* phase-shifted fringes, one only transmit the phase-demodulated signal as compressed sensing data.

## 1. Introduction

Optical phase-metrology has productively incorporated well known results from digital signal processing, stochastic processes and telecommunications [1-5]. In particular, we have applied the frequency transfer function (FTF) paradigm to the theory of phase-shifting algorithms (PSAs) [3]. The FTF is now widely used in phase-shifting algorithm (PSA) theory to investigate their signal-to-noise-ratio ($S/N$), their harmonic and detuning robustness [3]. Also stochastic processes theory applied to linear system has been used in PSAs to study the signal-to-noise-ratio ($S/N$) of the phase demodulation processes [1,3,4,5]. By plotting the absolute value of the FTF associated to a PSA one may gauge graphically its measuring noise, detuning and harmonic rejection [3].

Shannon information theory has not been applied to wavefront phase-metrology [4-11]. Many scientific and engineering disciplines, including optics, use Shannon theory to study the information processing behavior of their systems [9,10]. Shannon built on Nyquist [7] and Hartley [8] to develop a general theory for reliable information transmission/storage over noisy channels [6]. Nowadays this is simply known as Shannon information theory [4,10]. In this work we use Shannon information theory [6] to quantify the information storage of noisy fringe-patterns and fringe-data compression by phase-shifting algorithms (PSAs) [1-3].

In digital phase-metrology one wrongly equals the amount of raw fringe data-bits with information. One normally think that a fringe-image with $XxY$ pixels, and 8 bit/pixel, has $8(XxY)$ bits of information; however these are raw-data bits, not Shannon information bits. We show that noisy fringe patterns have much lower Shannon information than the capacity of their digitizing cameras. Quantifying the Shannon information storage of noisy fringe-images is important for fringe data compression for phase-measuring telemetry. For example, a spaceborne interferometer may take many phase-shifted fringe-images and send them back to earth for phase-demodulation. Or phase demodulate them within the spacecraft, and send just a single complex-valued demodulated signal as compressive sensing wave-data [13].

Knowing the Shannon information storage of noisy-fringes is also important because costly, high ($S/N$) CCD-cameras may not be necessary to digitize phase-modulated fringes for precision phase-metrology. That is because noisy fringe-images contains only a fraction of the information capacity of the CCD cameras which digitize them. Most bits/pixel information are wasted to speckle/electronic fringe-noise; for example an electronic speckle interferometry (ESPI) fringe-image, probably waste 7 out of 12 bits/pixel to noise. So a much cheaper, high resolution, low ($S/N$) camera, may be enough for ESPI metrology. In this work we show that poor ($S/N$) fringe-images may be compensated by large bandwidth $B$ fringes. For example, low ($S/N$) smartphone cameras, compensate this with high resolution images. The SCT allows us to think seriously about low ($S/N$), high resolution, mass-produced smartphone cameras for high-quality wave phase-metrology.

## 2. Mathematical theory of communication

Here we outline two important contributions on the mathematical theory of information by Claude E. Shannon: information-source entropy and channel-capacity [6]. Figure 1 shows the Shannon's abstraction of a telecommunication process [6]. The information source (the message) may be continuous or discrete data [4]. The message may then be amplitude or phase-modulated, digitally encoded/compressed, and finally transmitted. The radio signal then propagate towards the receiver. The communication channel is assumed bandlimited and noisy. The received signal is then decoded to obtain a reliable estimation of the original message. Signal decoding may consist on phase-demodulation, decompressing and error correction-algorithms [4]. We finally produce a recovered information which is a reliable copy the original message.

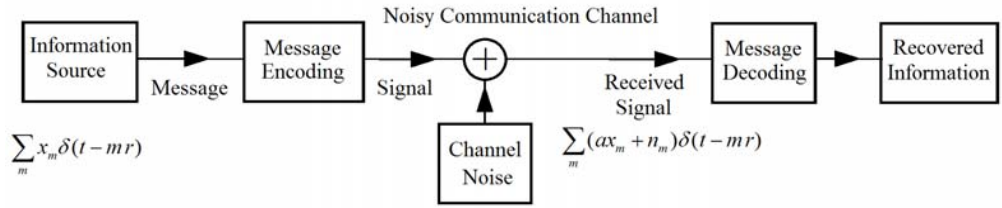

Fig. 1. Shannon abstraction of a memoryless noisy digital communication process in electrical engineering. The information source generate $r$ symbols/second [6].

Claude Shannon proved a formula for the channel capacity (the SCT) needed to transmit a message over a memoryless, noisy-channel with negligible information degradation [6]. In this paper [6], Shannon introduced the information-entropy and channel-capacity formulas for reliable transmission over a noisy additive white Gaussian noise (AWGN) channel [6].

*2.1 Shannon information entropy of a discrete information source*

Shannon information entropy $H$ of a discrete message consisting of a large sequence $\sum x_m \delta(t - m r)$ of symbols, drawn from $\{x_0, x_1, ..., x_{K-1}\}$, each sampled with probability $p(x_k) > 0$ is [6],

$$H = -\sum_{k=0}^{K-1} p(x_k) \log_2 [p(x_k)] \frac{\text{bits}}{\text{symbol}}; \quad \sum_{k=0}^{K-1} p(x_k) = 1.0 .. \quad (1)$$

In the average, a large data-sequence $\sum x_m \delta(t - m r)$ have an information entropy of $H$ bits/symbol. If these symbols $\{x_k\}$ are transmitted at a rate of $r$ symbols/second then, the information-rate is,

$$R = rH = -r \left\{ \sum_{k=0}^{K-1} p(x_k) \log_2 [p(x_k)] \right\} \frac{\text{bits}}{\text{second}} . \quad (2)$$

The prototypical example of a discrete source is the English language where the data are $\{A, B, C, ..., Z\}$, with $p(A) + ... + p(Z) = 1.0$, being $\{p(A), ..., p(Z)\}$ the probability of occurrence of a single letter in a text, then $H = -[p(A) \log_2 p(A) + ... + p(Z) \log_2 p(Z)]$.

*2.2 Shannon capacity theorem (SCT) for a bandlimited AWGN-channel*

The message $\sum x_m \delta(t - m r)$, is then transmitted over a AGWN-channel (see Fig. 1). The received signal is,

$$X(t) = \sum_m (a x_m + n_m) \delta(t - m r); \quad t \in [0, T]; \quad T \gg 1. \quad (3)$$

Where $a < 1.0$ is the channel attenuation. Shannon proved that the rate of information that a bandlimited AWGN-channel can transmit with negligible errors is [4-6,10],

$$C = B \log_2 \left(1 + \frac{S}{N}\right) \frac{\text{bits}}{\text{second}}; \quad S = a^2 E\{x_m^2\}; \quad N = E\{n_m^2\} .. \quad (4)$$

Where $B$ is the channel's bandwidth in cycles/second, $S$ is the received signal-power, $N$ is the channel noise-power, and $E\{\cdot\}$ is the ensemble average. This is the famous Shannon capacity theorem (SCT) for a bandlimited AWGN-channel [4-6,10-11]. The relation between the source information-rate $R$ and channel capacity $C$ for reliable communication is,

$$R > C \; ; Unreliable\ Communication \quad ; \quad . \tag{5}$$
$$R < C \; ; Negligible\ Error\ Communication.$$

Therefore in order to transmit reliably through an AWGN-channel the information-rate $R$ must be below $C$. For example, a typical AWGN telephone line with $B = 3\text{kHz}$ and $(S/N)=10^3$ has a capacity of $C = 3000\log_2(1+10^3) \approx 30,000$ bits/second.

The SCT also tell us that we can interchange ($S/N$) with bandwidth $B$ while keeping the channel capacity $C$ constant, as Table 1 shows,

Table 1. Channel capacity $C=\log_2[1+(S/N)]$ trade between ($S/N$) and bandwidth $B$

| Channel capacity $C$ (bits/sec) | Channel Bandwidth $B$ (Hz). | Signal-to-noise ratio ($S/N$) |
|---|---|---|
| 30,000 | 1,500 | 1,100,000 |
| 30,000 | 3,000 | 1,000 |
| 30,000 | 6,000 | 31 |
| 30,000 | 9,000 | 9 |
| 30,000 | 12,000 | 5 |

To keep $C$ constant, $B$ increases linearly while ($S/N$) decreases exponentially. That is, for low ($S/N$), one may keep $C$ constant by increasing $B$, and still have reliable communication; this is why spread-spectrum communication is so robust against thermal-noise, radio-jamming and low-power spacecraft communications [4]. A daily example of the use of the SCT is when one walks away from a Wi-Fi transmitter. Given that the hardware have not changed, the channel bandwidth $B$ remains the same. However at a large distance from the Wi-Fi transmitter the signal-power decreases. This reduces the data transfer rate $R$ towards our laptop because the channel encoding must be more redundant, less information-rate efficient.

*2.3 Shannon's capacity is a mathematical existence theorem*

Finally keep in mind that Shannon capacity theorem (SCT) is a theoretical upper-bound limit for reliable classical communication/storage of information independently of any engineering apparatus. The SCT is an existence theorem, it say nothing on how to implement it physically or algorithmically. That is, the SCT is a mathematical abstraction disembodied of any implementation, such as Einstein's $E=mc^2$ is not a receipt for building a nuclear reactor. In fact the SCT may be seen as a sphere-packing theorem in a $K$-dimensional Euclidian space [6,10,12]. In other words, no matter how advanced (non-quantum) digital-modem one may built, it cannot reliably transmit more classical information than $C$. This is like trying to obtain more energy than $E=mc^2$ from a rest mass, or trying to build a perpetual moving machine bypassing the second law of thermodynamics. Modern digital communication modems using efficient block error-correcting codes approaches the SCT limit, *i.e.* $R \to C$.

## 3. Shannon information storage of noisy fringe-patterns

Many signals that we communicate/store are continuous such as voice, video, or telemetry, where the discrete-data entropy $H$ cannot be used. The source information-rate $R=R(H,r)$ and channel-capacity $C=C(B,S,N)$ are both given in bits/second, however they use different parameters [6]. The entropy $H$, the rate $r$, and signal-power $S$ are information-source variables [6]. On the other hand, the bandwidth $B$, and AWGN-power $N$ are communication channel variables [6,4,10]. To estimate the source information-rate of a continuous-signal, one would need a continuous-density entropy. However a continuous-density entropy is not always well-defined [10]. In contrast, the bandwidth, the noise, and the signal-power of a noisy continuous-source are always well-defined. So it is better to have an information-rate formula for continuous-signals based on its bandwidth and signal-to-noise ratio, rather than on its continuous-density entropy [10,11]. Fortunately mathematician Andrei Kolmogorov proposed an information-rate formula for continuous, noisy information-sources, the $\varepsilon$-entropy [11,12]. That is, Kolmogorov $\varepsilon$-entropy does not depend on a continuous-density entropy which is good, because it avoids some theoretical problems [10,11,12].

In wave phase-metrology the message is the continuous phase-field which modulate the fringes; it could be an optical wavefront $W(x,y)$, or a solid $z = h(x,y)$ in fringe-projection profilometry [1,2]. The model for a continuous fringe-pattern is,

$$I(x,y) = a(x,y) + b(x,y)\cos\left[\frac{2\pi}{\lambda}W(x,y)\right] + n(x,y). \qquad (6)$$

Here $a(x,y)$, $b(x,y)$ are smooth functions, and $\lambda$ the illumination wavelength. The probability density $p(I)$ of these fringes is continuous. Shannon defined the differential entropy of a continuous random variable $X$ with density $p(x)$ as [6,10],

$$H = -\int p(x)\log_2[p(x)]dx; \qquad \int p(x)dx = 1.0. \qquad (7)$$

As in every example involving an integral, or even a density, we should include the statement *if it exist* [10]. It is easy to construct examples of random variables for which a density function does not exist or for which the above integral does not exist [10]. Instead of trying to estimate a continuous-density entropy $H$ of a fringe-image, we use the Kolmogorov's $\varepsilon$-entropy $H_\varepsilon(\xi)$. In 1956 Kolmogorov wrote [11]: According to the proper meaning of the word, the entropy of a signal with a continuous distribution is always infinite. If the continuous signals can, nevertheless, serve to transmit finite information, then it is only because they are always observed with bounded accuracy. Consequently, it is natural to define the appropriate "$\varepsilon$-entropy" $H_\varepsilon(\xi)$ of the object $\xi$ by giving the accuracy of observation $\varepsilon$. Shannon did this under the designation "rate of creating information with respect to a fidelity criterion" [11]. In our notation, the noiseless object is $\xi = a + b\cos[\varphi]$, which is observed with bounded-accuracy $\varepsilon = n$. Kolmogorov proved that $\bar{H}_\varepsilon(I)$ approximates Shannon's capacity $C$ [11, 12]. The information-rate $\bar{H}_\varepsilon(I)$ for $I(x,y)$ is [11],

$$\bar{H}_\varepsilon(I) = B_I \log_2(1 + \text{SNR}_K)\frac{\text{bits}}{\text{pixel}}; \qquad B_I \text{ in } \frac{\text{fringes}}{\text{pixel}}. \qquad (8)$$

Being $\text{SNR}_K$ the signal-to-noise ratio of $I(x,y)$, given by,

$$\text{SNR}_K = \frac{\iint\limits_{(x,y)\in\Omega} s^2(x,y)\,dxdy}{\iint\limits_{I(x,y)} n^2(x,y)\,dxdy} = \frac{\iint\limits_{(u,v)\in\Pi_B} |s(u,v)|^2\,dudv}{\iint\limits_{[-\pi,\pi]\times[-\pi,\pi]} |n(u,v)|^2\,dudv}; \quad s(x,y) = b\cos[\varphi]. \qquad (9)$$

Where $s(u,v)$, $n(u,v)$ are the spectra of $s(x,y)$, $n(x,y)$ respectively; $(x,y) \in \Omega$ is the region with well defined fringes; $(u,v) \in \Pi_B$ the fringes' spectral lobe; $B_I$ is the average bandwidth (see Fig. 2). Note that formally $\bar{H}_\varepsilon(I)$ equals $C$; but the difference is that now $B$ and $\text{SNR}_K$ refer to the noisy signal, not the noisy channel. Kolmogorov $\bar{H}_\varepsilon(I)$ cannot exceed the storage-capacity of the digitizing CCD-camera [11]. For example, for a 256 gray-levels per pixel camera, then $\bar{H}_\varepsilon(I) <$ 8 bits/pixel.

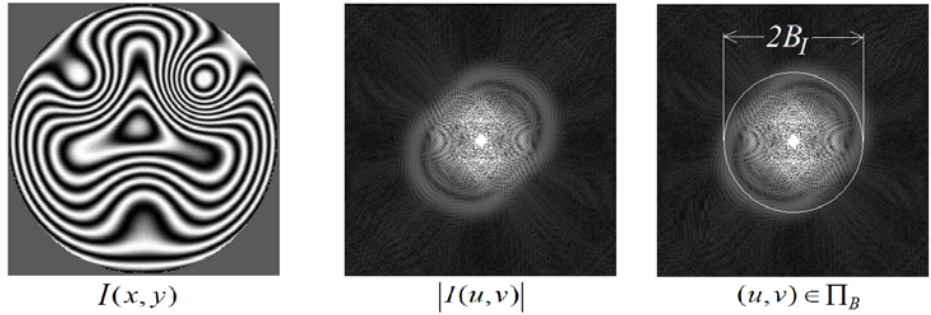

Fig. 2. A 8 bits/pixel quantized, noiseless simulated fringes $I(x,y)$. The spectrum $|I(u,v)|$ has a bandwidth $B_I$.

Let us consider the limits for $\bar{H}_\varepsilon(I)$ when the fringe's noise $n(x,y)$ tends to zero or infinite,

$$\lim_{n \to 0}\left[ B_I \log_2(1+\mathrm{SNR}_K) \right] = \infty; \qquad \lim_{n \to \infty}\left[ B_I \log_2(1+\mathrm{SNR}_K) \right] = 0. \tag{10}$$

The information-rate is infinite $\bar{H}_\varepsilon(I) = \infty$ for noiseless fringes, and zero $\bar{H}_\varepsilon(I) = 0$ for infinite noisy fringes; regardless of $B_I$. One might think that a constant-signal would hold little information; however if that signal is coded into a large number with say, $1 \times 10^8$ significant digits (above the noise $n$), then one may encode a large amount of information within that single number.

As mentioned, Shannon capacity $C$ (Eq. (3)) and Kolmogorov entropy-rate $\bar{H}_\varepsilon(I)$ (Eq. (7)) look identical, just substituting $(S/N)$ for $\mathrm{SNR}_K$. However in Shannon's capacity $N$ is a Gaussian stochastic processes [6], while the observation accuracy $\varepsilon$, in $\bar{H}_\varepsilon(I)$, is a deterministic error-signal [11,12]. This seems as an irrelevant theoretical nuance, but it has however practical consequences [6,10,11,12]. Lim and Franceschetti proved that $C$ and $\bar{H}_\varepsilon(I)$, are indeed "identical" under appropriate mathematical conditions [12].

## 4. Simulated Shannon information storage of fringe-images

Here we give two simple but illustrative examples of noisy fringe information storage. We use low resolution fringes to show the fringes at their Nyquist sampling rate of $B_I=0.5$ fringes/pixel. Figure 3 shows three 200x200-pixel, computer-generated noisy fringes phase-modulated by defocus and increasing $n(x,y)$.

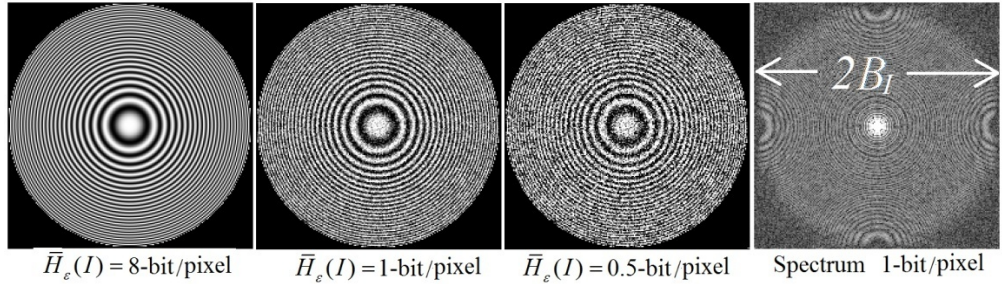

Fig. 3. Information storage-rate for fringes with $B_I=0.5$ fringes/pixel. The 8 bit/pixel fringes have $\mathrm{SNR}_K=70{,}000$ (about 8-bit quantizing noise). The noise is obvious by 1.0 bit/pixel ($\mathrm{SNR}_K=3.0$). The noisier fringes have 0.5 bits/pixel ($\mathrm{SNR}_K=1.0$).

We then decrease the SNR$_K$, until reaching $\bar{H}_\varepsilon(I)$ =0.5-bits/pixel. Figure 3 and Fig. 4 show the interplay between $B_I$ and SNR$_K$ stated by $\bar{H}_\varepsilon(I)$. Figure 4 shows a similar sequence of fringes but with reduced bandwidth $B_I$=0.125 fringes/pixel. In Fig. 4 the maximum fringe-frequency is one fourth of the Nyquist rate: $B_I$=0.25(0.5)=0.125 fringes/pixel, therefore the average information-rate $\bar{H}_\varepsilon(I)$ is reduced one fourth with respect to Fig. 3 for the same SNR$_k$.

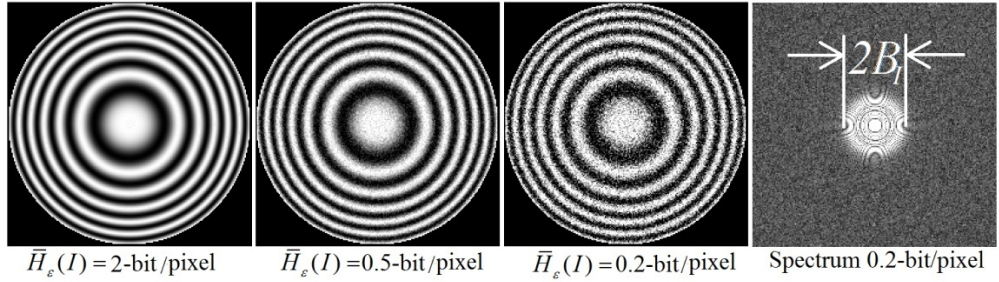

$\bar{H}_\varepsilon(I)=$ 2-bit/pixel   $\bar{H}_\varepsilon(I)=$ 0.5-bit/pixel   $\bar{H}_\varepsilon(I)=$ 0.2-bit/pixel   Spectrum 0.2-bit/pixel

Fig. 4. Average information-rate of fringe-images with $B_I$=0.125 fringes/pixel. The 2-bit/pixel fringes have SNR$_K$=70,000 (about 8-bits quantizing noise). The noisier fringes have 0.2 bits/pixel (SNR$_K$=2.03). The reduced spectrum of the noisier fringes is at far right.

The information-rate $\bar{H}_\varepsilon(I)$ is reduced by decreasing the SNR$_k$ until the noisier fringes at the far-right with 0.2-bit/pixel is obtained. Finally the spectrum of the noisiest-fringes with 0.2 bits/pixel is shown at the far right.

Remember that we are finding the low information-rate $\bar{H}_\varepsilon(I)$ of noisy fringes with respect to the CCD camera used to digitize them. Kolmogorov's $\bar{H}_\varepsilon(I)$ tell us nothing about how to program a compressing algorithm to store/transmit efficiently a large number of noisy fringe images.

## 5. Shannon information storage of experimental fringes

Here we are estimating the Shannon information content of experimental fringe-projection profilometry fringes. To estimate $\bar{H}_\varepsilon(I)$ for profilometry fringes, one possibility is to take two images: the noisy fringes, and the noisy background. The formalism for both images are [1],

$$I(x,y) = a(x,y) + b(x,y)\cos\left[\frac{2\pi}{p}\tan(\theta)h(x,y)\right] + n(x,y);$$
$$I_n(x,y) = a(x,y) + n(x,y).$$
(11)

Where $h(x,y)$ is the object being digitized; $p$ the period of the projected fringes and $\theta$ is the sensitivity angle between the projector and the (8 bits/pixel) CCD-camera [1].

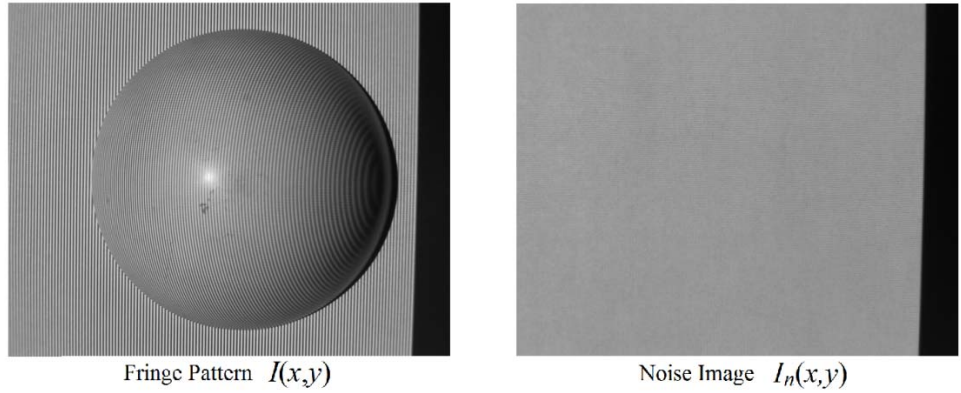
Fig. 5. High spatial frequency fringe-images (512x640, 8 bits/pixel) for obtaining an estimation of the information storage-rate for fringe-projection profilometry carrier-fringes.

White-light fringe-projection patterns have relatively high $SNR_K$. We then find the spectrum of the fringes $I(u,v) = F\{I(x,y)\}$, and define the indicator function for the signal-plus-noise region as $(u,v) \in SN$, and for the noise-only region as $(u,v) \in NR$.

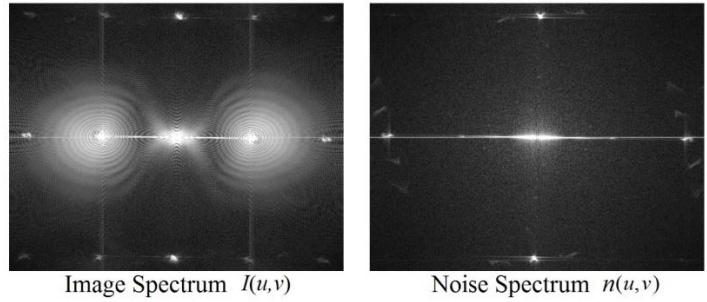
Fig. 6. Fourier spectrum corresponding to the fringes and noise (512x640 pixels).

Figure 7 shows the Fourier spectrum of the profilometry fringes and the Fourier spectrum of the background-image; both images defined in Eq. (11). Figure 7 shows the spatial filters to keep just the signal-lobes spectra within the region $(u,v) \in SN$.

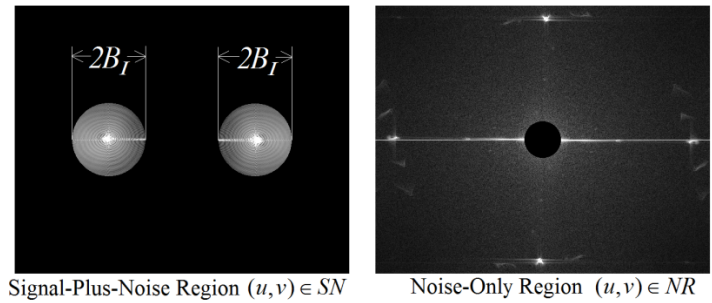
Fig. 7. Fourier filtered fringe-lobes ($B_I$=0.11), and high-pass filtered noise.

Next we show the demodulated wrapped-phase obtained by taken the right lobe of the signal-plus-noise region $(u,v) \in SN$ in Fig. 7. This was made to be sure that we have included all the phase signal bandwidth. This is shown in Fig. 8,

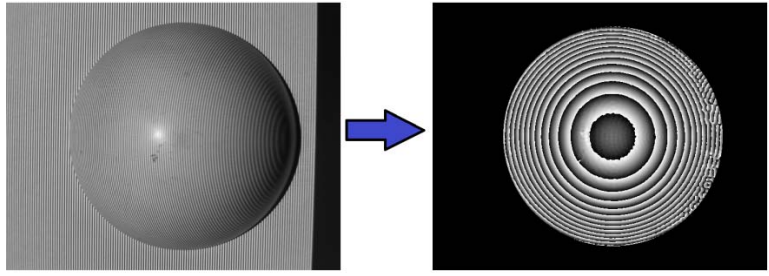

Fig. 8. The fringe-pattern and its wrapped demodulated phase (512x640 pixels). At the fringe self-occluding shadow the fringe-amplitude drops and the wrapped phase becomes noisier.

We then find the noise-power density within the noise-only region $(u,v) \in NR$ as,

$$\eta = \frac{1}{A_{NR}} \sum_{(u,v) \in NR} |a(u,v) + n(u,v)|^2. \tag{12}$$

Where $a(u,v) + n(u,v) = F\{a(x,y) + n(x,y)\}$ and $A_{NR}$ the area of $(u,v) \in NR$. We then assume that the noise-density within $(u,v) \in SN$ is also $\eta$ (Eq. (15)). Finding (S/N) as,

$$\text{SNR}_K = \frac{\sum_{(u,v) \in SN} \left(|I(u,v)|^2 - \eta\right)}{512(640)\eta} = 11.2. \tag{13}$$

We finally estimate the information-rate (with $B_I$=0.11, Fig. 7) of this fringe-pattern as,

$$\bar{H}_\varepsilon(I) = B_I \log_2\left(1 + \text{SNR}_K\right) = 0.41 \frac{\text{bits}}{\text{pixel}}. \tag{14}$$

The average bandwidth $B_I$ (Fig. 7) of the fringes may be estimated by looking at the spectrum the fringes $I(u,v) = F\{I(x,y)\}$. This profilometry fringe-image has in the average 0.41 bits/pixel, much less information than the 8 bits/pixel capacity of the CCD-camera. The information storage of most fringe patterns is usually well below the capacity of the digitizing CCD-camera. This means that if we want to transmit/store fringe-images one may effectively use compression image algorithms.

## 6. Real-life versus fringe-pattern images data compression

Real life images are highly redundant; have low information entropy. A real life photograph usually have smooth image regions, along with sparse noisy-like regions with detailed image features. Data compression algorithms may use local space-frequency wavelets expansion to reduce redundancy [13]. Image compression algorithms are broadly classified as lossy and lossless. In lossy compression the encoding/decoding algorithm decompress the original image not 100% accurately. However, subjective quality perception may tolerate large compression ratios without subjective degradation [10]. In lossless algorithms, decompressed images remains 100% accurate to the original. Therefore the most compressible images are usually smooth piece-wise continuous. On the other hand, one cannot compress a purely-random image, if we want to preserve exactly every pixel value; a white noise image is not compressible. Common image storage formats are: 8-bits/pixel Bit Maps (BMP); Joint Photographic Experts Group (JPEG); 8-bits/pixel Graphic Interchange Format (GIF); Tagged Image File Format (TIFF), and lossless Portable Network Graphics (PNG). These file formats have been designed to store real-life color images. These formats store noisy fringe-patterns as images containing more "information" than noiseless fringes. The noise is interpreted as fine image-details that the user wants to "preserve". On the contrary, noiseless fringes occupy less storage bytes; they are interpreted as smoother images (see Fig. 9).

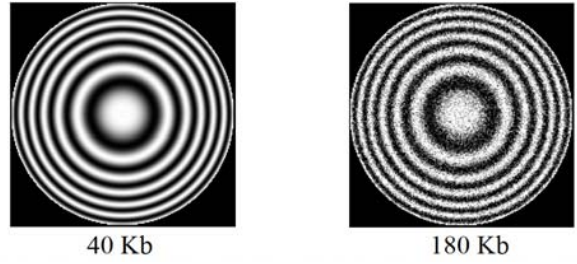

Fig. 9. Noiseless and noisy fringes with 200x200 pixel resolution and 8 bits/pixel. At right, the PNG file size is 40-Kbytes, while the noisy fringe image has a PNG file size of 180-Kbytes.

We have simulated two fringe patterns shown in Fig. 9. We have store them using lossless PNG format, and Fig. 9 shows the resulting PNG file sizes in kilobytes. The noisy image file is about four times bigger because the noise is interpreted as high frequency image details. The PNG format is designed to compress color real-life images for good visual perception, not phase-modulated noisy fringes. Fringe patterns have more mathematical structure than real-life images. So noisy fringe images are far more redundant and as a consequence, more compressible. The mathematical model for fringe patterns (Eq. (6)) allow us to distinguish between phase information and noise. Due to these considerations, real-life image compressors are not suitable for efficient transmission/storage of noisy fringe patterns; they were simply not designed for that purpose.

## 7. Fringe-data compression by phase-shifting algorithms (PSAs)

We have seen that noisy fringe patterns store much less Shannon information than the capacity of the CCD used to digitize them. In this section we show that phase-shifted fringes are even more information redundant. In remote optical/RADAR phase-metrology one may digitize $M$ phase-shifted fringes at a spacecraft (see Fig. 10). Then, one may downlink these $M$ fringes sequentially, and phase-demodulate them at the receiver. Or we may use a PSA to phase-demodulate them at the spacecraft, and downlink just the demodulated analytic-signal. A PSA may then be regarded as compressive-sensing, where we are "sensing" not real-valued fringes, but less noisy complex-valued wave-data [13]. So PSA compression of phase-shifted fringes is paramount in remote phase-shifting metrology where communications are limited due to large distances and few watts for radio transmission. A PSA then packs the information contained in $M$ phase-shifted images into a single phase-demodulated complex-valued image.

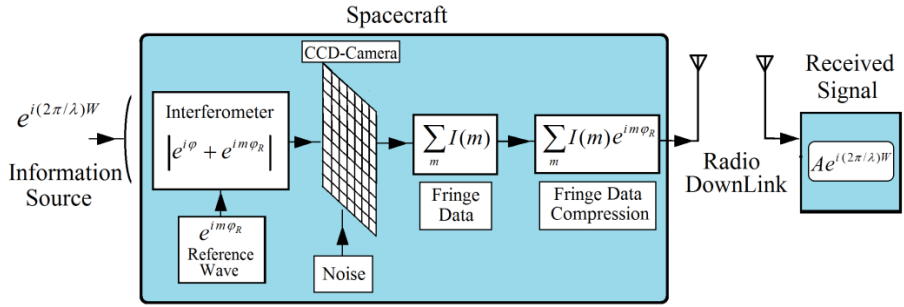

Fig. 10. Schematic of a spacecraft transmitting high-information, complex-valued wave-data from a set of $M$ phase-shifted fringe-patterns.

To quantify the increase of information-rate by phase-shifting algorithms (PSAs), consider a sequence of $M$ phase-shifted noisy interferograms,

$$I(x,y,t) = \sum_{m=0}^{M-1}\left\{a + b\cos\left[\frac{2\pi}{\lambda}W + m\omega_0\right] + n_m\right\}\delta(t-m); \quad \omega_0 = \frac{2\pi}{M}. \tag{15}$$

Here some (*x,y*) coordinates were omitted for clarity. We may phase-demodulate these fringes by a PSA whose impulse response is [3],

$$h(t) = \sum_{m=0}^{M-1} c_m \, \delta(t-m). \tag{16}$$

Being $c_m$ complex-valued. The Fourier transform of $h(t)$ ($H(\omega) = F[h(t)]$) must have at least the following zeroes [3],

$$H(-\omega_0) = H(0) = 0; \quad \text{and} \quad H(\omega_0) \neq 0. \tag{17}$$

Obtaining the estimated quadrature analytic signal as [3],

$$Ae^{i\varphi(x,y)} = I(x,y,t) * h(t)\big|_{t=M-1} = \sum_{m=1}^{M-1} c_m \, I(x,y,m). \tag{18}$$

A PSA increases the (*S/N*) of the analytic-signal with respect to the fringes (Eq. (15)) by [3],

$$G_{S/N} = \frac{|H(\omega_0)|^2}{\frac{1}{2\pi}\int_{-\pi}^{\pi}|H(\omega)|^2 d\omega}. \tag{19}$$

The highest $G_{S/N}$ is obtained by the least-squares PSA, where $G_{S/N} = M$ [3]. Then the $SNR_K$ of the noisy fringes increases to $M(SNR_K)$ for the complex-valued phase-demodulated signal $Ae^{i\varphi(x,y)}$. Then the information-rate contained in the analytic signal $Ae^{i\varphi(x,y)}$ is,

$$\bar{H}_\varepsilon(A;M) = B_I \log_2\left(1 + M \, SNR_K\right) \frac{\text{bits}}{\text{pixel}}. \tag{20}$$

The information of $Ae^{i\varphi(x,y)}$ increases logarithmically with *M*. An interesting question is: how many phase-shifted fringes double the information content of $\bar{H}_\varepsilon(A;M)$, or,

$$B_I \log_2\left(1 + M \, SNR_K\right) = 2\left\{B_I \log_2\left(1 + SNR_K\right)\right\}. \tag{21}$$

Solving for *M* one obtains,

$$M = 2 + SNR_K. \tag{22}$$

This is shown in Fig. 11,

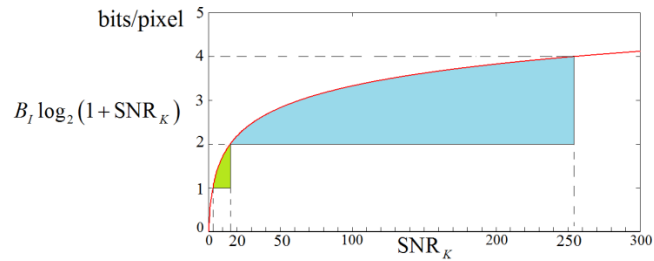

Fig. 11. Information storage as a function of $SNR_K$, with $B_I$=0.5 fringes/pixel. At the green-zone, the spatial information-rate increases almost linearly from 1 to 2 bits/pixel. In contrast in the blue-zone, this rate increases logarithmically from 2 to 4 bits/pixel.

For example with $\text{SNR}_K=1$ one would need $M=3$ phase-shifted fringes to double the information-rate of $Ae^{i\varphi(x,y)}$. In contrast for $\text{SNR}_K=20$, one would need $M=22$ phase-shifted fringes to double the information-rate of $Ae^{i\varphi(x,y)}$ from 2 to 4 bits/pixel. An example of this is shown in Fig. 12 and Fig. 13 (for $M=12$).

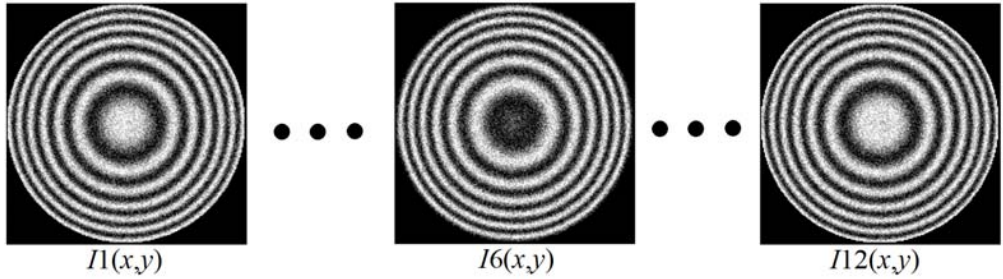

Fig. 12. Four out-of 12 phase-shifted fringes degraded with AWGN. These fringes have an average information-rate of 0.4bits/pixel ($B_I$=0.125fringes/pixel, $\text{SNR}_K$=8.1).

The $M=12$ phase-shifted fringes in Fig. 12 were phase-demodulated using a least-squares PSA and the resulting real and imaginary images of the analytic signal are shown in Fig. 13,

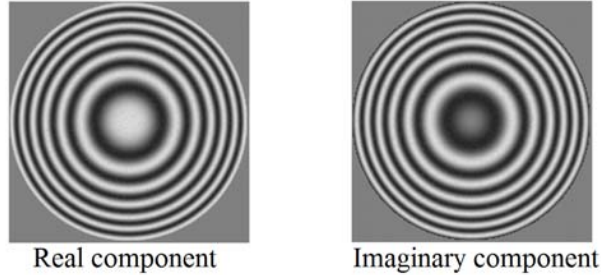

Fig. 13. Real and imaginary parts of the phase-demodulated signal from the 12 noisy phase-shifted fringes in Fig. 12. The information-rate has increased about twice, from 0.4 bits/pixel to 0.82 bits/pixel ($B_I$=0.12fringes/pixel, $\text{SNR}_K$=97.5, $M$=12).

As conclusion, a PSA regarded as fringe-image compressor packs into a single analytic-signal $Ae^{i\varphi(x,y)}$ the phase information stored in $M$ phase-shifted fringes. In other words, instead of transmitting/storing $M$ phase-shifted fringes, one may send/store a single, higher-information signal $Ae^{i\varphi(x,y)}$, or wrapped-phase $\arg[Ae^{i\varphi(x,y)}]$. Therefore a PSA may be regarded as an efficient compression algorithm for transmission or storage of large blocks of phase-shifted fringe-data. This is paramount for efficient transmission of optical/RADAR interference fringes from remote phase-metrology sites where the channel is severely limited by large distances (high attenuation) and low power radio link. Also a PSA may be regarded as a compressive-sensing algorithm where the collected data is complex-valued with lower-noise and as consequence higher information rate [13].

## 8. Summary

We have shown that noisy fringe-images in wave phase-metrology contain much less Shannon information than the capacity of the CCD camera used to digitize them. For example, it does not make sense to store electronic speckle-pattern interferometry (ESPI) fringes using a costly 14 bits/pixel cooled CCD-camera. That is because most data bits (not information bits) are wasted to fringe-noise. ESPI noisy fringes have very-low spatial information-rate. Shannon information may be used as cost-efficiency criteria for choosing a CCD camera according to the class of fringes that we are dealing with. For example, mass-produced smartphone cameras may do just fine in many optical-metrology cases. Smartphone cameras have low signal-to-noise ratio, but this is compensated with high image-pixel resolution; 12 megapixels or more is nowadays available. So smartphone cameras with moderate (*S/N*) and very high resolution images, may be efficiently used for remote phase-wave telemetry applications.

We showed that low-noise fringe-images may be obtained from white-light, fringe-projection profilometry. Here we have estimated that an experimental carrier-fringe profilometry image may contain less than 0.5 bits/pixel of Shannon information (Figs. 5-6). With these small information-rates one realizes that there is plenty of room for noisy fringe-data compression for efficient transmission or storage. We have seen that PSAs may be regarded as fringe-data compression algorithms. Using PSAs, all Shannon information contained in *M* phase-shifted fringes may be compressed into a single analytic signal. In spaceborne phase-metrology for example, one may use PSAs to compress large blocks of *M* phase-shifted fringes for efficient downlink. Finally we saw that PSAs compress noisy fringe-data information by increasing the $\text{SNR}_K$ of the phase-demodulated complex-signal.